\documentclass[aps,prl,showpacs,reprint]{revtex4-1}
\usepackage{amsmath,amssymb,graphicx,subfigure,xcolor}
\usepackage[english]{babel}
\usepackage[utf8]{inputenc}
\usepackage{epstopdf}

\newcommand{\te}{\theta}

\newcommand{\tte}{\tilde{\theta}}

\newcommand{\tP}{\tilde{P}}
\newcommand{\J}{J_{jk}}

\newcommand{\pa}{\partial}

\newcommand{\lan}{\langle}
\newcommand{\ran}{\rangle}

\newcommand{\om}{\omega}

\newcommand{\ph}{\varphi}

\newcommand{\si}{\sigma}

\newcommand{\cD}{\mathcal{D}}

\begin{document}

\title{Nature of the Volcano Transition in the Fully Disordered Kuramoto Model}
\author{Axel Prüser}
\email{axel.prueser@uol.de}
\author{Sebastian Rosmej}
\author{Andreas Engel}
\affiliation{Carl von Ossietzky University Oldenburg, Institut für Physik, D26111 Oldenburg, Germany}

\begin{abstract}
Randomly coupled phase oscillators may synchronize into disordered patterns of collective motion. We analyze this transition in a large, fully connected Kuramoto model with symmetric but otherwise independent random interactions. Using the dynamical cavity method, we reduce the dynamics to a stochastic single-oscillator problem with self-consistent correlation and response functions that we study analytically and numerically. We clarify the nature of the volcano transition and elucidate its relation to the existence of an oscillator glass phase.
\end{abstract}

\maketitle

Synchronization is ubiquitous in nature and science~\cite{Strobook}. Periodic degrees of freedom with different individual frequencies tend to lock into phase coherent motion when weakly coupled to one another. Physics, chemistry, biology, engineering, and even sociology abound in examples~\cite{PiRoKu,Strobook,PiRo}. Quantitative studies flourished after 1984 when Kuramoto introduced a mean-field model of phase oscillators with random frequencies of spread $\si$ globally coupled with strength $J$~\cite{Kubook}. For small ratios $J/\si$ the time evolution of the oscillators remains incoherent, whereas for large $J/\si$ they continue to change in unison. More precisely, the incoherent state remains stable up to a threshold value of $J/\si$~\cite{StMi91} followed by a sharp transition into an ordered state with gradually increasing degree of synchronization~\cite{Kubook}. Over the years, the original Kuramoto model with identical couplings between all oscillators has been investigated thoroughly, and a wealth of information on it is by now available~\cite{Stro00,Acebron,PiRo}.

Much less is known about the inhomogeneous situation in which the oscillators are connected by couplings $J_{ij}$ of different strength and sign~\cite{Acebron}. The asymmetric case with no relation between $J_{ij}$ and $J_{ji}$ is dominated by chaotic dynamics with little tendency to synchronization~\cite{SoCrSo88}. For symmetric couplings, $J_{ij}=J_{ji}$, synchronization prevails and, due to frustration, new phenomena emerge. Some of these have been discussed in~\cite{BoPeRu92} and in a series of recent papers by Strogatz and collaborators investigating different types of symmetric but disordered coupling matrices $J_{ij}$~\cite{HoStro11a,HoStro11b,KloLiStro14,OtStro18}. Partial synchronization, mixed phases, antiphase synchronization, traveling-wave states, and other patterns were found.

The system of central interest in this field, however, is the analog of the Sherrington-Kirkpatrick (SK) model of spin glasses~\cite{SK} with symmetric but otherwise independent couplings drawn from a Gaussian distribution. It was studied numerically in pioneering work by Daido more than 30 years ago~\cite{Daido92}. At $J/\si\simeq 1.3$, he found a peculiar ``volcano'' transition in the distribution of the order parameter that gave rise to speculations about the existence of an oscillator glass state and remained mysterious ever since. 

So far, no analytical theory is available for this system. An attempt made in~\cite{StRa98} remained essentially restricted to the case of asymmetric couplings. The methods developed in~\cite{OtStro18} when applied to the SK setting result in an exponential number of order parameters rendering the approach unfeasible~\cite{OtStro18,PaGa23}. 

In the present Letter, we analyze the Kuramoto model of phase oscillators with normally distributed frequencies and random interactions of SK type. Using the dynamic cavity approach, we reduce the dynamics of the $N$-oscillator system to a self-consistent stochastic differential equation for a single oscillator that we study analytically and numerically. We reproduce the volcano transition found by Daido, provide a physical picture of its nature, and discuss its relation to a possible oscillator glass phase. 


The model is defined by the set of equations 
\begin{equation}\label{basiceq}
 \pa_t\te_j(t)=\om_j+h_j(t)+\sum_k \J \sin\big(\te_k(t)-\te_j(t)\big)
\end{equation} 
for the time evolution of the phases $\te_j$ of $N$ oscillators $j= 1,...,N$. The frequencies $\om_j$ are drawn at random from a normal distribution with zero mean and unit variance, and the initial conditions $\te_j(0)=\te_j^{(0)}$ derive from a flat distribution over $[-\pi,\pi)$. The couplings are symmetric, $\J=J_{kj}$, and otherwise independent Gaussian random variables with zero mean and variance $J^2/N$. The local fields $h_j(t)$ are  introduced solely to define the susceptibilities  $\chi_{jk}(t,t'):=\delta \te_j(t)/\delta h_k(t')$ and will be put equal to zero otherwise. We are interested in the long-time dynamics of the system in the thermodynamic limit $N\to\infty$. 

Because of the global coupling, the model is of mean-field type and may be reduced to an effective, self-consistent single-site problem. To accomplish this reduction, we employ the dynamical cavity method that rests on the assumption of stochastic stability of the thermodynamic limit~\cite{MPV}. We consider system~\eqref{basiceq} for one particular realization of frequencies, initial conditions, and couplings and {\em add} one new oscillator, $j=0$, with phase $\te_0(t)$ and new quenched random variables $\om_0,\,\te_0^{(0)}$, and $J_{0k}$. 

The presence of the new oscillator induces slight deviations $\te_j\to\tte_j=\te_j+\delta\te_j$ in the dynamics of the previous ones. Since $J_{0k}=O(1/\sqrt{N})$, they are small and may be treated in linear response:
\begin{equation}\label{defdelte}
 \delta\te_j(t)=\sum_l\!\int_0^t\!\! dt'\, 
     \chi_{jl}(t,t') J_{l0}\sin\big(\te_0(t')-\te_l(t')\big).
\end{equation}
To leading order in $N$, the equation of motion of the new oscillator then assumes the form 
\begin{widetext}
 \begin{equation}\label{eqte0a}
 \pa_t\te_0(t)= \om_0+h_0(t)+\sum_k J_{0k} \sin\big(\te_k(t)-\te_0(t)\big)
    +\sum_{k,l} J_{0k} J_{l0}\int_0^t\!\! dt'\,\chi_{kl}(t,t')\cos\big(\te_k(t)-\te_0(t)\big)\sin\big(\te_0(t')-\te_l(t')\big).
 \end{equation}
\end{widetext}
The new couplings $J_{0k}$ are statistically independent of the $\te_i(t),\, i=1,...,N$, and, therefore, the third and fourth term on the rhs of this equation are large sums of independent random contributions with finite first and second moments. By the central limit theorem, they are, therefore, Gaussian random functions and may be characterized by their respective first and second moments. As it turns out, the third term gives rise to a Gaussian noise  with zero mean, whereas the fourth term is characterized by negligible fluctuations for $N\to\infty$ and may be replaced by its average~\cite{sm}. Denoting by $\lan\dots\ran$ the combined average over all couplings (including the new ones) as well as over the frequencies and initial conditions for $i\geq 1$, Eq.~\eqref{eqte0a} acquires the form~\cite{sm}
\begin{align}\nonumber
  \pa_t\te_0(t)&=\om_0+ h_0(t)+J \cos\te_0(t)\,\xi_2(t)-J\sin\te_0(t)\,\xi_1(t)\\\label{eqte0b}
        &\qquad\quad-J^2\!\!\int_0^t\!\! dt'\,R(t,t')\sin\big(\te_0(t)-\te_0(t')\big).
\end{align} 
Here $\xi_i,\,i=1,2$, are two independent Gaussian noise sources with
\begin{equation}\label{statxi}
 \lan\xi_i(t)\ran\equiv 0,\qquad \lan\xi_i(t)\xi_j(t')\ran=\delta_{ij}\,C(t,t').
\end{equation}
Moreover, 
\begin{align}\label{defCprel}
 C(t,t')&:=\frac{1}{2N}\sum_j\lan\cos\big(\te_j(t)-\te_j(t')\big)\ran\\\label{defRprel}
 R(t,t')&:=\frac{1}{2N}\sum_j\lan\chi_{jj}(t,t')\cos\big(\te_j(t)-\te_j(t')\big)\ran.
\end{align} 
The noise terms derive from the disordered crosstalk of the $\te_j$ on $\te_0$ and are, hence, proportional to $J$. The response term stems from the feedback of the polarization of the $\te_j$ due to $\te_0$ on $\te_0$ itself and is  proportional to $J^2$. 

Consequently, the phase $\te_0(t)$ of the new oscillator obeys a Langevin equation with multiplicative colored noise and memory kernel with correlation and response functions determined by the dynamics of the original $N$-oscillator system in the {\em absence} of $\te_0$. We now close the argument by stipulating that in the averaged $(N+1)$-oscillator system $\te_0$ should in no way be special. Therefore the average in~\eqref{defCprel} and~\eqref{defRprel} may be replaced by one over the dynamics of $\te_0(t)$ itself. 

In this way we end up with our self-consistent single-oscillator problem defined by~\eqref{eqte0b} and~\eqref{statxi} together with 
\begin{align}\label{defC}
 C(t,t')&:=\frac{1}{2}\lan\cos\big(\te_0(t)-\te_0(t')\big)\ran\\\label{defR}
 R(t,t')&:=\frac{1}{2}\lan\chi_{00}(t,t')\cos\big(\te_0(t)-\te_0(t')\big)\ran\\\label{defchi}
 \chi_{00}(t,t')&:=\frac{\delta\te_0(t)}{\delta h_0(t')}.
\end{align}
The average $\lan\dots\ran$ is now over the realizations of the $\xi_i$ as well as over $\om_0$ and $\te_0^{(0)}$. In the following, the by now superfluous index~$0$ will be omitted. For large values of $t$ and $t'$, the functions $C, R$, and $\chi$ will depend only on $\tau:=t-t'$.  Equations equivalent to~\eqref{eqte0b},~\eqref{statxi}, and ~\eqref{defC}-\eqref{defchi} may also be derived with the help of the Martin-Siggia-Rose formalism~\cite{MSR} as applied to systems with quenched disorder~\cite{DeDo78,SoZi82,SoCr18}. 

The self-consistent stochastic single-site problem derived above represents a difficult system of equations. Analytical progress is possible by perturbation theory in $J$ which we worked out up to fourth order~\cite{sm}:
\begin{align}\label{Cpert}
 C(\tau)&=C_0(\tau)+J^2 C_2(\tau)+J^4 C_4(\tau)+O(J^6)\\\label{Rpert}
 R(\tau)&=R_0(\tau)+J^2 R_2(\tau)+J^4 R_4(\tau)+O(J^6).
\end{align}
The complete expressions are rather long, and we quote here only the asymptotic behavior for large $\tau$:
\begin{equation}\label{asymp}
 \begin{split}
 C_0(\tau)&\sim R_0(\tau)\sim\frac{1}{2}e^{-\tau^2/2}\\
 C_2(\tau)&\sim 3\,R_2(\tau)\sim\frac{3\sqrt{\pi}}{16}\,\tau\,e^{-\tau^2/4}\\
 C_4(\tau)&\sim 13\, R_4(\tau)\sim\frac{13\pi\sqrt{3}}{864}\,\tau^2\,e^{-\tau^2/6}.
 \end{split}
\end{equation}
As can be seen, the decay of both correlation and response functions becomes slower with increasing order, and one may speculate about the general asymptotic behavior
\begin{equation*}
 C_{2n}(\tau),\, R_{2n}(\tau)\sim \tau^n\,e^{-\frac{\tau^2}{2(n+1)}}.
\end{equation*}

\begin{figure}[t]
 \includegraphics[width=.4\textwidth]{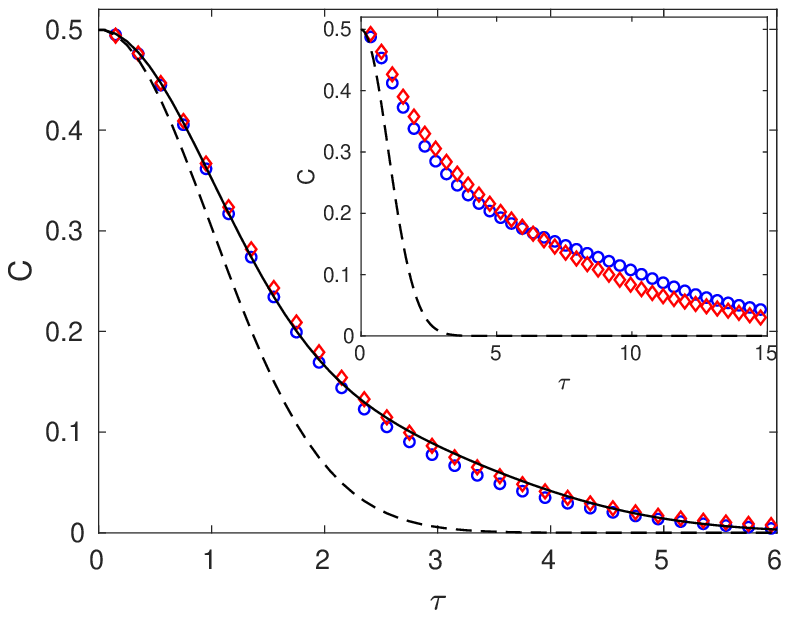}
 \caption{Correlation function $C(\tau)$ as obtained via~\eqref{defCprel} from simulations of an $N=10^3$ oscillator network averaged over 50 disorder realizations (red diamonds) and as resulting from~\eqref{defC} and $5\cdot 10^4$ trajectories of the single-oscillator dynamics (blue circles). Statistical errors are smaller than the symbol size. The main figure is for $J=0.80$, the inset for $J=1.75$. Also shown are the lowest-order perturbative result $C_0(\tau)$ (black dashed line) and, for $J=0.80$,  the fourth-order result~\eqref{Cpert} (black full line).}
 \label{fig1}
\end{figure}

The single-oscillator dynamics may also be studied by numerical simulations~\cite{EiOp,Roy}. Starting with initial guesses for $C(t,t')$ and $R(t,t')$ and generating a family of trajectories $\{\te(t)\}$ according to~\eqref{eqte0b}, one determines refined approximations for $C(t,t')$ and $R(t,t')$ via~\eqref{defC} and~\eqref{defR} and iterates until convergence is reached~\cite{sm}. In this procedure, the functional derivative $\delta/\delta h(t')$ of Eq.~\eqref{eqte0b} yields an equation of motion for $\chi(t,t')$ which is solved in parallel with the one for $\te(t)$~\cite{Roy}. This numerical analysis is rather different from and hopefully complementary to the direct simulation of the original $N$-oscillator problem~\eqref{basiceq} as performed by Daido. In particular, no finite-size effects have to be taken into account: The solution of~\eqref{eqte0b} is expected to display the statistical properties of a typical oscillator in an $N=\infty$ system~\cite{EiOp}. Finite-size effects are a notorious nuisance in the simulation of glassy systems ~\cite{BY,Daido00}, and their absence is quite advantageous.

To validate our approach, we display in Fig.~\ref{fig1} the correlation function $C(\tau)$ as obtained from the original $N$-oscillator system and~\eqref{defCprel}, the one determined self-consistently according to~\eqref{defC}, and analytical results from perturbation theory. For $J=0.80$, i.e., below the volcano transition, the agreement is very good. At $J=1.75$, well above $J_c\simeq 1.3$, both simulations show a markedly increased correlation range as can be inferred from the comparison with $C_0(\tau)$. Unfortunately, the validity of our fourth-order perturbation theory does not extend up to $J=1.75$.


We now turn to the analysis of the volcano transition. As order parameters that are able to signal synchronization into {\em disordered} phase patterns, Daido introduced the so-called complex local fields~\cite{Daido92}: 
\begin{equation}\label{defpsi}
 \Psi_j(t):=r_j(t)\,e^{i\phi_j(t)}:=\frac{1}{J}\sum_k\J\,e^{i\te_k(t)}.
\end{equation}
With their help, Eq.~\eqref{basiceq} may be rewritten as 
\begin{equation*}
 \pa_t\te_j(t)=\om_j+h_j(t)+Jr_j(t)\sin\big(\phi_j(t)-\te_j(t)\big).
\end{equation*}
From this form, it is apparent that the distribution $P(r)$ of the moduli $r_j$ is crucial for the overall tendency of the system to synchronize. The $\phi_j$ merely determine the specific pattern on which the phases lock. In fact, Daido detected the volcano transition in histograms of 
\begin{equation*}
 \tP(r):=\frac{P(r)}{2\pi r}
\end{equation*}
compiled from time series $r_j(t)$ for different $j$. For small coupling strength, $\tP$ has its maximum at $r=0$. With increasing $J$, the height of this maximum shrinks, and at $J=J_c$ it turns into a minimum with another maximum developing at nonzero value of $r$. 

In view of~\eqref{defpsi}, the order parameter for the newly added oscillator is given by 
\begin{equation}\label{defpsi0}
 \Psi(t):=r(t)\,e^{i\phi(t)}:=\frac{1}{J}\sum_j J_{0j}\, e^{i\tte_j(t)}.
\end{equation}
Using $\tte_j=\te_j+\delta\te_j$ together with~\eqref{defdelte}, it may be split into two parts:
\begin{align}\nonumber
 \Psi(t)&=\frac{1}{J}\sum_j J_{0j}\, e^{i\te_j(t)}+\frac{i}{J}\sum_j J_{0j}\, e^{i\te_j(t)}\delta\te_j(t)+\dots\\
        &=:\xi(t)+\zeta(t).\label{decompsi}
\end{align}
Here, $\xi=\xi_1+i\xi_2$ is a complex Gaussian noise defined by~\eqref{statxi}. Arguments similar to those that carried us from~\eqref{eqte0a} to~\eqref{eqte0b} reveal that the fluctuations of $\zeta$ as induced by the randomness in the couplings and the $\te_j, j\geq 1$ are negligible for $N\to\infty$. As a result, we get~\cite{sm}
\begin{equation}\label{defzeta}
 \zeta(t):=\rho(t)\,e^{i\ph(t)}:=J\int_0^t\!\! dt'\,R(t,t')\,e^{i\te_0(t')},
\end{equation} 
with
\begin{equation}\label{defrho}
 \rho(t)=J \left|\int_0^t\!\! dt'\,R(t,t')\,e^{i\te_0(t')}\right|.
\end{equation} 

The decomposition~\eqref{decompsi} of $\Psi$ now clearly reveals the mechanism behind the volcano transition. For $J=0$, the order parameter is equal to $\xi$, and, therefore, it is a Gaussian random variable with zero mean. Accordingly, the maximum of $\tP(r)$ is at $r=0$. With increasing $J$, the Gaussian distribution for $\Psi$ is systematically shifted such that its maximum coincides with $\zeta$. This reduces $\tP(0)$, and for sufficiently large $|\zeta|$ the maximum of $\tP$ at $r=0$ turns into a minimum; see also Fig.~3 in~\cite{sm}. Note that the crucial prefactor $J$ in~\eqref{defzeta} results from the different $J$ scaling of noise and response terms in~\eqref{eqte0b}.

More quantitatively, we have 
\begin{equation}\label{defpofr}
 P(r(t)):=\left\lan \int\!\!\cD\xi_1(\cdot)\cD\xi_2(\cdot)\,\delta\big(r(t)-\left|\Psi(t)\right|\big)\right\ran,
\end{equation}
where the outer average is over $\om$ and $\te^{(0)}$. Further analytical progress is now hampered by the fact that due to its dependence on $\te_0(t')$ the order parameter $\Psi(t)$ is a complicated functional of the $\xi_i(\cdot)$. Both our perturbative results and our numerical simulations, however, indicate~\cite{sm} that typically $\rho(t)$ is very near to its upper bound 
\begin{equation}\label{defrhoup}
 \rho_u:=J\int_0^t\!\! dt'\,\left|R(t,t')\,e^{i\te_0(t')}\right|\to J\int_0^\infty\!\! d\tau\,|R(\tau)|,
\end{equation} 
where in the last step the limit $t\to\infty$ was taken. 

Using $\rho_u$ instead of $\rho$ in~\eqref{defpofr}, no dependence on $\om$ and $\te^{(0)}$ remains, and, moreover, the path integrals over the noise histories simplify to Gaussian averages over $\xi_i(t)$. Performing these, we end up with~\cite{sm}
\begin{equation}\label{respofr}
 \tP(r)\simeq \tP_u(r)=\frac{1}{\pi}\,e^{-r^2-\rho_u^2}\,I_0(2r\rho_u),
\end{equation} 
where $I_0$ denotes a modified Bessel function~\cite{AS}. From 
\begin{equation*}
 \frac{\pa^2 \tP_u}{\pa r^2}\Bigg|_{r=0}=\frac{2}{\pi}\,e^{-\rho_u^2}(\rho_u^2-1)
\end{equation*}
the corresponding approximation $J_{cu}$ for $J_c$ is implicitly given by $\rho_u(J_{cu})=1$. Using the lowest-order perturbative result for $R(\tau)$~\cite{sm}, we find
\begin{equation*}
 J_{cu}=\frac{4}{\sqrt{2\pi}}\simeq 1.6.
\end{equation*}

\begin{figure}[ht]
 \includegraphics[width=.4\textwidth]{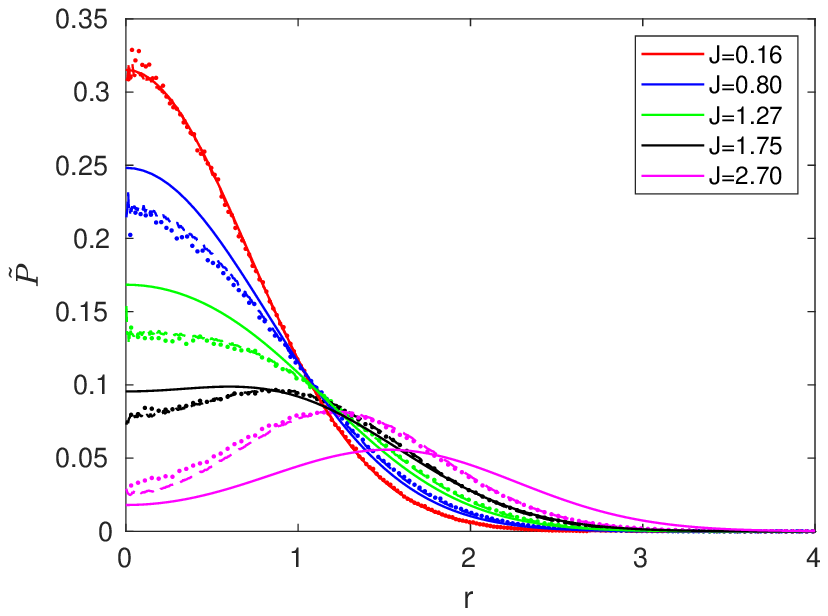}
 \caption{Comparison between $\tP_u(r)$ as determined by~\eqref{respofr} (full lines) with numerical simulations of $10^3$ trajectories of the effective single-oscillator problem~\eqref{eqte0b} (dotted lines) and of the original $N$-oscillator problem~\eqref{basiceq} with $N=10^3$ averaged over 20 disorder realizations (dashed lines).}
 \label{fig2}
\end{figure}

Figure ~\ref{fig2} compares~\eqref{respofr} with numerical simulations. For small $J$, the agreement is fine. With increasing $J$, the nonlocal correlations between $\zeta(t)$ and $\xi(t')$ that were neglected in the derivation of~\eqref{respofr} become more and more important. Nevertheless, the qualitative agreement between approximate analytical theory and numerical simulations remains valid. In particular, the volcano transition is clearly described by~\eqref{respofr}. The histograms from both numerical simulations are consistent with those of Daido~\cite{Daido92} and, hence, also reproduce his result  $J_c\simeq 1.3$, which is somewhat smaller than $J_{cu}$.

We finally discuss the relation of our findings to the possible existence of an oscillator glass phase. Our  perturbation results~\eqref{asymp} for the correlation and response functions show a tendency to long-range autocorrelations and slowly decaying memory kernel with increasing values of $J$. Although this may be taken as precursors of a freezing transition, these asymptotics give no hint on a power law decay that is generally believed to characterize a glass phase \cite{rem}. Moreover, in the analysis of~\eqref{eqte0b} for large $J$, we do not find any indication for a persistent part in the correlation function similar to the case of the SK spin glass~\cite{SoZi82}. This is consistent with the phase diffusion observed by Daido for very long times~\cite{Daido92} and may originate in the peculiar symmetries of the model that show up also in the linear stability analysis of the incoherent state~\cite{StMi91}. Also, the mechanism behind the volcano transition described by the decomposition~\eqref{decompsi} of the order parameter is rather general, and it seems likely that it may be found in several models~\cite{OtStro18,PaGa23} irrespective of whether these possess a glass phase or not. Finally, from the investigation of so-called ``noise-induced'' transitions, it is well known that it may be misleading to link a thermodynamic transition to the topological change of a probability distribution~\cite{BrScSc82}. In line with~\cite{OtStro18,PaGa23}, we, therefore, believe that the volcano transition found in~\cite{Daido92} does not immediately lead into a glass phase. 

In conclusion, we have provided an intuitive as well as an analytical understanding of the volcano transition characteristic for synchronization processes in disordered networks of coupled phase oscillators. A  key ingredient is the decomposition \eqref{decompsi} of the local order parameter in a random and a systematic part. Our methods should be relevant for systems beyond networks of simple phase oscillators. Examples include optomechanical arrays~\cite{HLQKM}, chemical systems~\cite{TRTSE}, and colonies of bacteria~\cite{PSRDTH} as well as the very recently discussed networks of Lohe~\cite{LJSK} and Stuart-Landau oscillators~\cite{PB}.

Stimulating discussions with Harald Engel, Joachim Krug, Alexander Hartmann, Satya Majumdar, Peter Reimann, Markus Schmitt, and Peter Young are gratefully acknowledged.

\bibliographystyle{apsrev4-1} 
\bibliography{paper} 

\begin{thebibliography}{35}
\bibitem [{rem()}]{rem}%
  \BibitemOpen
  \href@noop {} {}\bibinfo {note} {This may, of course, also be due to our
  weak-coupling perturbation theory failing to describe the frozen
  phase.}\BibitemShut {Stop}%
\end{thebibliography}

\end{document}


\begin{center}
 {\large\bf Supplementary Material for}\\[2ex]
 {\large\bf ``Nature of the volcano transition in the fully disordered Kuramoto model''}\\[2ex]
  
  Axel Pr\"user, Sebastian Rosmej, and Andreas Engel\\[1ex]

 Carl von Ossietzky University Oldenburg, Institut für Physik, D26111 Oldenburg, Germany
\end{center}

\vspace*{.5cm}

\section{Basic equations}

We consider the system of equations
\begin{equation}\label{basiceq}
 \pa_t\te_j(t)=\om_j+h_j(t)+\sum_k \J \sin\big(\te_k(t)-\te_j(t)\big)\spa{with}
 \te_j(0)=\te_j^{(0)}
\end{equation} 
for $j=1,..,N$. The parameters in~\eqref{basiceq} are random variables with distributions
\begin{align}\label{Pom}
 P\big(\{\om_j\}\big)&=\prod_j \frac{1}{\sqrt{2\pi}}\,\exp\left(-\frac{\om_j^2}{2}\right)\\\label{PJ}
 P\big(\{\J\}\big)&=\prod_{j<k} \frac{1}{\sqrt{2\pi J^2/N}}\,\exp\left(-N\frac{\J^2}{2J^2}\right),
  \quad J_{kj}=\J\\\label{Pte0}
 P\big(\{\te_j^{(0)}\}\big)&=\prod_j \frac{1}{2\pi}\quad,\qquad\qquad -\pi\leq\te_j^{(0)}< \pi\quad .
\end{align}
It will be important to understand how $\te_j(t)$ is modified by a small change in the r.h.s. of~\eqref{basiceq}. For 
\begin{equation*}
 h_k(t) \to h_k(t)+\delta h_k(t)
\end{equation*}
with small $\delta h_k$ we may use linear response theory stipulating 
\begin{equation}\label{linres}
 \te_j(t)\to\te_j(t)+\delta\te_j(t)\spa{with} 
  \delta\te_j(t) =\sum_k\int_0^t\!\! dt'\, \chi_{jk}(t,t')\delta h_k(t'),
\end{equation} 
where we have defined the susceptibility
\begin{equation}\label{defchi}
 \chi_{jk}(t,t'):=\frac{\delta \te_j(t)}{\delta h_k(t')}.
\end{equation} 
This definition was the only reason to include the field $h_j(t)$ in \eqref{basiceq}. Whenever solutions $\te_j(t)$ of \eqref{basiceq} are considered it is understood that they correspond to $h_j(t)\equiv 0$ for all $j$.

Taking the functional derivative of~\eqref{basiceq} we get an equation of motion for $\chi$,
\begin{align}\nonumber
 \pa_t \chi_{jl}(t,t')=\pa_t \frac{\delta \te_j(t)}{\delta h_l(t')}
      &=\frac{\delta h_j(t)}{\delta h_l(t')} +\sum_k \J \cos\big(\te_k(t)-\te_j(t)\big)
      \left[\frac{\delta \te_k(t)}{\delta h_l(t')}-\frac{\delta \te_j(t)}{\delta h_l(t')}\right]\\
 &=\delta_{jl}\,\delta(t-t')+\sum_k \J \cos\big(\te_k(t)-\te_j(t)\big)
      \left[\chi_{kl}(t,t')-\chi_{jl}(t,t')\right]\label{eqmchi},
\end{align}
with initial condition $\chi(t,t')=0$ for $t<t'$.

We will later assume 
\begin{equation}\label{chioffdiag}
 \chi_{jl}(t,t')=\mathcal{O}\left(1/\sqrt{N}\right)\spa{for} j\neq l
\end{equation} 
whereas the $\chi_{jj}$ are of order one. This assumption is self-consistent with~\eqref{eqmchi} because for $j\neq l$
\begin{equation}\label{h1}
 \pa_t \chi_{jl}(t,t')=\sum_k \J \cos\big(\te_k(t)-\te_j(t)\big)
      \left[\chi_{kl}(t,t')-\chi_{jl}(t,t')\right].
\end{equation}
The parameters $\J$ are random and of order $1/\sqrt{N}$. Hence, the r.h.s. of this equation consists of one term ($k=l$) of order $1/\sqrt{N}$ and a sum of $(N-2)$ {\em random} terms of order $1/N$ which is also $\mathcal{O}(1/\sqrt{N})$. Therefore, the r.h.s. of~\eqref{h1} is of order $1/\sqrt{N}$ and with $\chi_{jk}(t,t')=0$ for $t<t'$ due to causality the nondiagonal susceptibilities will stay of this order  for $t>t'$ as well.


\section{Symmetries}

Eqs.\eqref{basiceq} -\eqref{Pte0} exhibit certain symmetries that simplify the calculations. To begin with we introduce the abbreviations
\begin{equation}\label{defabbrev}
 \te:=\te(t),\quad \te':=\te(t'),\quad \bte(t,t'):=\te-\te',\quad \hte(t,t'):=\te+\te'.
\end{equation} 
First, an overall shift in initial conditions, 
\begin{equation}\label{shifttheta}
 \te_j^{(0)} \to \te_j^{(0)}+\Theta\qquad \forall j
\end{equation}
results in a {\em permanent} overall shift 
\begin{equation}\label{shifte}
 \te_j(t) \to \te_j(t)+\Theta\qquad \forall j\spa{and} \forall t>0.
\end{equation}
This is due to the fact that the sine functions in~\eqref{basiceq} all have phase {\em differences} as arguments. 
Therefore, and since~\eqref{Pte0} gives the same probability to all initial conditions $\te_j^{(0)}$, the set of trajectories $\{\te_j(t)\}$ has the same probability (with respect to $\te_j^{(0)}$) as the set $\{\te_j(t)+\Theta\}$. This in turn implies for any functional $f$ and all $\Theta$
\begin{equation*}
 \lan f[\{\te_j(\cdot)\}]\ran_{\te^{(0)}}=\lan f[\{\te_j(\cdot)+\Theta\}]\ran_{\te^{(0)}}, 
\end{equation*}
where $\lan\dots\ran_{\te^{(0)}}$ denotes the sole average with respect to $\te_j^{(0)}$. 

The particular choice $f[\{\te_j(\cdot)\}]=e^{i(\te_k+\te_k')}=e^{i\hte_k(t,t')}$ yields 
\begin{equation*}
 \lan e^{i\hte_k(t,t')}\ran_{\te^{(0)}}=\lan e^{i(\hte_k(t,t')+2\Theta)}\ran_{\te^{(0)}}
 =e^{2i\Theta}\lan e^{i\hte_k(t,t')}\ran_{\te^{(0)}}
\end{equation*}
which can be satisfied for all $\Theta$ only if 
\begin{equation*}
 \lan e^{i\hte_k(t,t')}\ran_{\te^{(0)}}\equiv 0\qquad \forall k.
\end{equation*}
The real and imaginary part of this equation are 
\begin{equation}\label{avhte}
 \lan \cos\hte_k \ran_{\te^{(0)}}\equiv 0\spa{and} \lan \sin\hte_k \ran_{\te^{(0)}}\equiv 0.
\end{equation}
Since the shift~\eqref{shifte} does not influence the susceptibilities we also have
\begin{equation}\label{avchite}
 \lan \chi_{kk}\cos\hte_k \ran_{\te^{(0)}}\equiv 0\spa{and} \lan \chi_{kk}\sin\hte_k \ran_{\te^{(0)}}\equiv 0.
\end{equation} 

Second, if $\te_j(t)$ are solutions of~\eqref{basiceq} then $\tte_j(t):=-\te_j(t)$ are solutions of 
\begin{equation*}
 \pa_t\tte_j(t)=-\om_j+\tilde{h}_j(t)+\sum_k \J \sin\big(\tte_k(t)-\tte_j(t)\big)\spa{with}
 \tte_j(0)=-\te_j^{(0)} \quad\text{and}\quad \tilde{h}_j(t)=-h_j(t).
\end{equation*} 
Since both $P(\om)$ and $P(\te^{(0)})$ are even functions of their respective arguments the trajectories $\{\te_j(t)\}$ hence have the same probability (with respect to $\om_j$ and $\te_j^{(0)}$) as the trajectories $\{-\te_j(t)\}$. Consequently,
\begin{equation*}
 \lan f[\{\te_j(t)\}] \ran_{\om,\te^{(0)}}\equiv 0
\end{equation*}
for any odd functional 
\begin{equation*}
 f[\{\te_j(t)\}]=-f[\{-\te_j(t)\}],
\end{equation*}
where $\lan\dots\ran_{\om,\te^{(0)}}$ denotes the average over all $\om_j$ and $\te^{(0)}_j$.
In particular 
\begin{equation}\label{avsin}
 \lan \sin\bte_j\ran _{\om,\te^{(0)}}\equiv 0 \qquad \forall j.
\end{equation} 
Moreover, since
\begin{equation*}
 \tilde{\chi}_{kl}(t,t'):=\frac{\delta \tte_k(t)}{\delta \tilde{h}_l(t')}
                         =\frac{\delta (-\te_k(t))}{\delta (-h_l(t'))}
                         =\frac{\delta \te_k(t)}{\delta h_l(t')}=\chi_{kl}(t,t'),
\end{equation*}
we also have
\begin{equation}\label{avchisin}
 \lan \chi_{jj}\sin\bte_j\ran _{\om,\te^{(0)}}\equiv 0.
\end{equation} 


\section{Adding a new oscillator}

Assume that the equations~\eqref{basiceq} have been solved for $j=1,..,N$ such that the $\te_j(t)$ are known functions for all choices of $\J,\om_j,\te_j^{(0)}$ and $h_j(t)$. We now add another oscillator $\te_0$ with new and {\em independently} drawn parameters $J_{j0}=J_{0j}, \om_0$, and $\te_0^{(0)}$ to the system. Its equation of motion reads
\begin{equation}\label{eq0}
 \pa_t\te_0(t)=\om_0+h_0(t)+\sum_k J_{0k} \sin\big(\tte_k(t)-\te_0(t)\big)\spa{and}
 \te_0(0)=\te_0^{(0)}.
\end{equation} 
The functions $\tte_k(t)$ differ slightly from $\te_k(t)$ because there is an additional term in their equations of motion due to the presence of $\te_0$:
\begin{equation*}
 \pa_t\tte_j(t)=\om_j+h_j(t)+J_{j0}\sin\big(\te_0(t)-\tte_j(t)\big)+\sum_k \J \sin\big(\tte_k(t)-\tte_j(t)\big).
\end{equation*} 
Since $J_{0k}=\mathcal{O}(1/\sqrt{N})$ the additional term is small and we may use~\eqref{linres} to find 
\begin{equation}\label{defdeltate}
 \tte_j(t)-\te_j(t)=:\delta\te_j(t)=\sum_l\int_0^t\!\! dt'\, 
     \chi_{jl}(t,t') J_{l0}\sin\big(\te_0(t')-\te_l(t')\big).
\end{equation} 
Higher order corrections due to the replacement $\tte_l \to \te_l$ on the r.h.s. of this equation can be  neglected.

Eq.~\eqref{eq0} therefore assumes the form 
\begin{align}\nonumber
 \pa_t\te_0(t)&=\om_0+h_0(t)+\sum_k J_{0k} \sin\big(\te_k(t)-\te_0(t)\big)
      +\sum_k J_{0k} \cos\big(\te_k(t)-\te_0(t)\big)
       \sum_l\int_0^t\!\! dt'\,\chi_{kl}(t,t') J_{l0}\sin\big(\te_0(t')-\te_l(t')\big) \\\label{h3}
    &=\om_0+h_0(t)+\sum_k J_{0k} \sin\big(\te_k(t)-\te_0(t)\big)
    +\sum_{k,l} J_{0k} J_{l0}\int_0^t\!\! dt'\,\chi_{kl}(t,t')\cos\big(\te_k(t)-\te_0(t)\big)\sin\big(\te_0(t')-\te_l(t')\big).
\end{align}
The third and the fourth term in this equation are random functions of time and we proceed by determining their statistical properties in the limit $N\to\infty$.
The third term describes the influence of the {\em unmodified} oscillators $\te_1, \dots, \te_N$ on the dynamics of $\te_0$. It can be written in the form
\begin{align}\nonumber
 \sum_k J_{0k} \sin\big(\te_k(t)-\te_0(t)\big)
  &=\cos\te_0(t)\sum_k J_{0k} \sin\te_k(t)-\sin\te_0(t)\sum_k J_{0k} \cos\te_k(t)\\\label{defxi}
  &=:J\cos\te_0(t)\,\xi_2(t)-J\sin\te_0(t)\,\xi_1(t).
\end{align}

Since the new couplings $J_{0k}$ are {\em independent} of all parameters determining the dynamics of the $\te_k(t)$ both $\xi_1$ and $\xi_2$ are large sums of independent random variables with finite first and second moment. Hence both are Gaussian random functions fixed by their average and correlations. 
Denoting by $\lan\dots\ran$ the combined average over $\J$ for $j,k=0,\dots, N$ and $\om_j,\te_j^{(0)}$ for $j=1,\dots, N$ we find
\begin{equation}\label{avxi}
 \lan\xi_1(t)\ran=\frac{1}{J}\sum_k \ub{\lan J_{0k}\ran}_{0} \lan\cos\te_k(t)\ran\equiv 0 \spa{and}
 \lan\xi_2(t)\ran=\frac{1}{J}\sum_k \ub{\lan J_{0k}\ran}_{0} \lan\sin\te_k(t)\ran\equiv 0
\end{equation} 
as well as
\begin{align*}
 \lan\xi_1(t)\xi_1(t')\ran&=\frac{1}{J^2}\sum_{j,k}\ub{\lan J_{0j}J_{0k}\ran}_{\frac{J^2}{N}\delta_{jk}}
  \lan\cos\te_j\cos\te_k'\ran
 =\frac{1}{N}\sum_j\lan\cos\te_j\cos\te_j'\ran
 =\frac{1}{2N}\sum_j\lan\cos\bte_j+\cos\hte_j\ran=\frac{1}{2N}\sum_j\lan\cos\bte_j\ran\\
 \lan\xi_2(t)\xi_2(t')\ran
&=\frac{1}{J^2}\sum_{j,k}\ub{\lan J_{0j}J_{0k}\ran}_{\frac{J^2}{N}\delta_{jk}}
  \lan\sin\te_j\sin\te_k'\ran
 =\frac{1}{N}\sum_j\lan\sin\te_j\sin\te_j'\ran
 =\frac{1}{2N}\sum_j\lan\cos\bte_j-\cos\hte_j\ran=\frac{1}{2N}\sum_j\lan\cos\bte_j\ran,
\end{align*} 
where we have used~\eqref{avhte}.
Since both $\xi_1$ and $\xi_2$ depend on $J_{0k}$ there may also be a cross-correlation. However, using~\eqref{avhte} and~\eqref{avsin} we find 
\begin{equation}\label{crosscorxi}
 \lan\xi_1(t)\xi_2(t')\ran=\frac{1}{J^2}\sum_{j,k}\ub{\lan J_{0j}J_{0k}\ran}_{\frac{J^2}{N}\delta_{jk}}
  \lan\cos\te_j\sin\te_k'\ran
 =\frac{1}{N}\sum_j\lan\cos\te_j\sin\te_j'\ran
 =\frac{1}{2N}\sum_j\lan\sin\hte_j-\sin\bte_j\ran\equiv 0.
\end{equation}
The influence of the {\em undisturbed} $\te_j(t)$ on the time-evolution of $\te_0$ can therefore be modelled by two uncorrelated Gaussian noise sources $\xi_1$ and $\xi_2$ the statistical properties of which are determined by the dynamics of the $\te_j(t), j\geq 1$ according to eqs.~\eqref{avxi} and 
\begin{equation}\label{corxi}
  \lan\xi_1(t)\xi_1(t')\ran= \lan\xi_2(t)\xi_2(t')\ran=C(t,t'):=\frac{1}{2N}\sum_j\lan\cos\bte_j(t,t')\ran.
\end{equation} 

The forth term in~\eqref{h3} can be written in the form $\int_0^t\! dt' R_\mathrm{tot}(t,t')$ where
\begin{equation}\label{defRlong}
 R_\mathrm{tot}(t,t'):=\sum_{k,l} J_{0k} J_{l0}\chi_{kl}(t,t') \cos\big(\te_k(t)-\te_0(t)\big)\sin\big(\te_0(t')-\te_l(t')\big),
\end{equation} 
describes the influence of changes induced by the presence of $\te_0$ on the $\te_j(t)$ back on the dynamics of $\te_0$ itself. It naturally gives rise to a response term. Due to the independence between new couplings and old phases also $R_\mathrm{tot}(t,t')$ is a random Gaussian function and we aim at the determination of its average and correlation function. 
First, 
\begin{align*}
 \cos\big(\te_k-\te_0\big)\sin\big(\te'_0-\te'_l\big)
 &=\frac{1}{2}\big[\sin(\te_k-\te_0+\te'_0-\te'_l)-\sin(\te_k-\te_0-\te'_0+\te'_l)\big]\\
 &=\frac{1}{2}\big[\sin(\te_k-\te'_l-\bte_0)-\sin(\te_k+\te'_l-\hte_0)\big]\\
 &=\frac{1}{2}\big[\cos\bte_0\sin(\te_k-\te'_l)-\sin\bte_0\cos(\te_k-\te'_l)
                  -\cos\hte_0\sin(\te_k+\te'_l) +\sin\hte_0\cos(\te_k+\te'_l)\big]
\end{align*} 
Accordingly, $R_\mathrm{tot}$ is a sum of four terms,
\begin{equation}\label{defRtot}
 R_\mathrm{tot}(t,t')=J^2\big[R_1(t,t')\cos\bte_0-R_2(t,t')\sin\bte_0-R_3(t,t')\cos\hte_0+R_4(t,t')\sin\hte_0\big],
\end{equation}
where
\begin{equation*}
 \begin{split}
  R_1(t,t')&:=\frac{1}{2J^2}\sum_{k,l} J_{0k} J_{l0}\chi_{kl}(t,t') \sin(\te_k-\te'_l),\qquad
  R_2(t,t'):=\frac{1}{2J^2}\sum_{k,l} J_{0k} J_{l0}\chi_{kl}(t,t') \cos(\te_k-\te'_l),\\
  R_3(t,t')&:=\frac{1}{2J^2}\sum_{k,l} J_{0k} J_{l0}\chi_{kl}(t,t') \sin(\te_k+\te'_l),\qquad
  R_4(t,t'):=\frac{1}{2J^2}\sum_{k,l} J_{0k} J_{l0}\chi_{kl}(t,t') \cos(\te_k+\te'_l).
 \end{split}
\end{equation*}
Let us first analyze the second one. We split it into the contribution from $k=l$ and the rest:
\begin{equation}\label{statR2a}
  R_2(t,t')=:\tR_2(t,t')+r_2(t,t')=\frac{1}{2J^2}\sum_k J_{0k}J_{k0}\chi_{kk}(t,t') \cos\bte_k(t,t') +
 \frac{1}{2J^2}\sum_{(k,l)} J_{0k} J_{l0}\chi_{kl}(t,t') \cos(\te_k-\te'_l).
\end{equation} 
Both $\tR_2$ and $r_2$ are Gaussian random functions. More precisely, we have
\begin{align*}
 \lan \tR_2(t,t')\ran&=\frac{1}{2J^2}\sum_k \lan J_{0k}J_{k0}\ran \lan\chi_{kk}(t,t') \cos\bte_k(t,t')\ran
   =\frac{1}{2N}\sum_k \lan\chi_{kk}(t,t') \cos\bte_k(t,t')\ran=\mathcal{O}(1)\\
 \lan \tR_2(t,t')\tR_2(s,s')\ran
      &=\frac{1}{4J^4}\sum_{k,l} \ub{\lan J_{0k}J_{k0}J_{0l}J_{l0}\ran}_{\frac{J^4}{N^2}\big(1+2\delta_{kl}\big)}
            \lan\chi_{kk}(t,t') \cos\bte_k(t,t')\,\chi_{ll}(s,s') \cos\bte_l(s,s')\ran\\
  &=\frac{1}{4N^2}\sum_{k,l} \lan\chi_{kk}(t,t')\cos\bte_k(t,t')\,\chi_{ll}(s,s') \cos\bte_l(s,s')\ran\\
  &\qquad\qquad+\frac{1}{2N^2}\sum_k\lan\chi_{kk}(t,t') \cos\bte_k(t,t')\,\chi_{kk}(s,s') \cos\bte_k(s,s')\ran.
\end{align*}
The second term in the last line is $\mathcal{O}(1/N)$ and may be neglected. We therefore find
\begin{align*}
 \lan \tR_2&(t,t')\tR_2(s,s')\ran-\lan \tR_2(t,t')\ran\lan \tR_2(s,s')\ran\\
 &=\frac{1}{4N^2}\sum_{k,l} \Big[\lan\chi_{kk}(t,t')\cos\bte_k(t,t')\,\chi_{ll}(s,s') \cos\bte_l(s,s')\ran
  -\lan\chi_{kk}(t,t') \cos\bte_k(t,t')\ran\lan\chi_{ll}(s,s')\cos\bte_l(s,s')\ran\Big].
\end{align*}
Assuming the correlations under the sum to be of order $1/N$ 
the fluctuations in $\tR_2(t,t')$ are of order $1/\sqrt{N}$ and for $N\to\infty$ it may be replaced by its average
\begin{equation}
 \tR_2(t,t')=\frac{1}{2N}\sum_k \lan\chi_{kk}(t,t') \cos\bte_k(t,t')\ran+\mathcal{O}(1/\sqrt{N}).
\end{equation} 
For $r_2(t,t')$ we get
\begin{align*}
 \lan r_2(t,t')\ran&\equiv 0\\
 \lan r_2(t,t')r_2(s,s')\ran&=\frac{1}{4J^4}\sum_{(k,l)} \sum_{(m,n)} 
   \ub{\lan J_{0k} J_{l0}J_{0m} J_{n0}\ran}_{\frac{J^4}{N^2}(\delta_{km}\delta_{ln}+\delta_{kn}\delta_{lm})}
   \big\lan\chi_{kl}(t,t') \cos\big(\te_k(t)-\te_l(t')\big)\,\chi_{mn}(s,s') \cos\big(\te_m(s)-\te_n(s')\big)\big\ran\\
 &=\frac{1}{4N^2}\sum_{(k,l)}\big\lan\chi_{kl}(t,t')\chi_{kl}(s,s')
    \cos\big(\te_k(t)-\te_l(t')\big)\cos\big(\te_k(s)-\te_l(s')\big)\\
  &\qquad\qquad\qquad +\chi_{kl}(t,t')\chi_{lk}(s,s')
    \cos\big(\te_k(t)-\te_l(t')\big)\cos\big(\te_l(s)-\te_k(s')\big) \big\ran.
\end{align*}
Assuming now $\chi_{ij}=\mathcal{O}(1/\sqrt{N})$ for $i\neq j$ we find also for $r_2(t,t')$ fluctuations of order $1/\sqrt{N}$ which may be neglected relative to the average of $\tR_2(t,t')$ for $N\to \infty$. Note also 
\begin{equation*}
 \lan \tR_2(t,t')r_2(s,s')\ran\equiv 0
\end{equation*}
due to the independence of the $J_{0j}$. 
Hence, the off-diagonal terms in the $(k,l)$-sum do not contribute at all to $R_2(t,t')$ and we end up with
\begin{equation}\label{defR2}
 R_2(t,t')=\frac{1}{2N}\sum_k \lan\chi_{kk}(t,t') \cos\bte_k(t,t')\ran.
\end{equation} 
For $R_1$ we get by similar means 
\begin{equation}\label{defR1}
 R_1(t,t')=\frac{1}{2N}\sum_k \lan\chi_{kk}(t,t') \sin\bte_k(t,t')\ran\equiv 0,
\end{equation} 
where the last identity follows from~\eqref{avchisin}.
The other two terms reduce to 
\begin{equation*}
 R_3(t,t')=\frac{1}{2N}\sum_k \lan\chi_{kk}(t,t') \sin\hte_k(t,t')\ran\spa{and}
 R_4(t,t')=\frac{1}{2N}\sum_k \lan\chi_{kk}(t,t') \cos\hte_k(t,t')\ran
\end{equation*}
respectively, which are both identically zero due to~\eqref{avchite}. Hence, only $R_2(t,t')$ survives in the response term \eqref{defRtot}. Denoting $R_2(t,t')$ simply by $R(t,t')$ we therefore end up with.
\begin{equation*}
R_\mathrm{tot}(t,t')=-J^2 R(t,t')\sin\bte_0(t,t').
\end{equation*}

Plugging all these results into~\eqref{h3} we get eq.~(4) of the main text:
\begin{equation}\label{LEsc1}
 \pa_t\te_0(t)=\om_0+h_0(t)+J \cos\te_0(t)\,\xi_2(t)-J\sin\te_0(t)\,\xi_1(t)
        -J^2\int_0^t\!\! dt'\,R(t,t')\sin\bte_0(t,t').
\end{equation} 
The ensuing equation of motion for $\chi_{00}(t,t')$ then is  
\begin{align}\nonumber
\pa_t\chi_{00}(t,t')=\frac{\delta\pa_t\te_0(t)}{\delta h(t')}\Big|_{h(t)\equiv 0}
   &=\delta(t-t')-J\chi_{00}(t,t')\big(\sin\te_0(t)\,\xi_2(t)+\cos\te_0(t)\,\xi_1(t)\big)\\\label{eomchi}
   &\qquad\qquad-J^2\int_0^t\!\!dt''\,R(t,t'')\cos\bte_0(t,t'')\big(\chi_{00}(t,t')-\chi_{00}(t'',t')\big)
\end{align} 
with initial condition 
\begin{equation*}
 \chi_{00}(t,t')=0\spa{for} t<t'.
\end{equation*}
From now on we omit the index $0$ to label the new oscillator.


\section{Perturbation theory}

As discussed in the main text the noise and response terms in the central eq.~(4) are proportional to $J$ and $J^2$, respectively. This invites a perturbative treatment for small values of $J$ which we performed up to fourth order. We sketch here the general setup of the perturbation expansion as well as the calculations of the zeroth and second order expressions. The fourth order analysis is analogous but much more tedious. 

With the perturbation ansatz
\begin{equation}\label{pertte}
 \te(t)=\te_0(t)+J\te_1(t)+J^2\te_2(t)+J^3\te_3(t)+J^4\te_4(t)+\cO(J^5)
\end{equation} 
we get 
\begin{align}\nonumber
 \cos\te\,\xi_2-\sin\te\,\xi_1&=\big(\cos\te_0\,\xi_2-\sin\te_0\,\xi_1\big)
      -\big(\sin\te_0\,\xi_2+\cos\te_0\,\xi_1\big)\big(J\te_1+J^2\te_2+J^3\te_3\big)\\\label{h5}
 &-\frac{1}{2}\big(\cos\te_0\,\xi_2-\sin\te_0\,\xi_1\big)\big(J\te_1+J^2\te_2\big)^2
  +\frac{1}{6}\big(\sin\te_0\,\xi_2+\cos\te_0\,\xi_1\big)J^3\te_1^3+\cO(J^4).
\end{align}
Note that the index $0$ now denotes the zeroth order of perturbation theory. It is convenient to introduce the abbreviations
\begin{equation}\label{defphpsi}
 \ph(t):=\cos\te_0(t)\,\xi_2(t)-\sin\te_0(t)\,\xi_1(t)\spa{and}
 \psi(t):=-\sin\te_0(t)\,\xi_2(t)-\cos\te_0(t)\,\xi_1(t).
\end{equation}
Both are Gaussian random functions with zero mean,
\begin{equation*}
 \lan\ph(t)\ran_\xi\equiv\lan\psi(t)\ran_\xi\equiv 0,
\end{equation*}
and correlations 
\begin{equation}\label{statphpsi}
  \lan \ph(t)\ph(t')\ran_\xi=\lan \psi(t)\psi(t')\ran_\xi=C(t,t')\cos\bte_0(t,t')\quad\text{and}\quad
  \lan \ph(t)\psi(t')\ran_\xi=-\lan \psi(t)\ph(t')\ran_\xi=C(t,t')\sin\bte_0(t,t').
\end{equation} 
With their help~\eqref{h5} can be rewritten in the form
\begin{align*}
 \cos\te\,\xi_2-\sin\te\,\xi_1
  &=\ph+J\psi\,\te_1+J^2\big(\psi\,\te_2-\frac{1}{2}\ph\,\te_1^2\big)
    +J^3(\psi\,\te_3-\ph\,\te_1\te_2- \frac{1}{6}\psi\,\te_1^3\big)  +\cO(J^4).
\end{align*}
We also expand $C(t,t')$ and $R(t,t')$ in powers of $J$:
\begin{align}
 C(t,t')&=\frac{1}{2}\lan\cos\bte(t,t')\ran=C_0(t,t')+JC_1(t,t')+J^2C_2(t,t')+
               J^3C_3(t,t')+J^4C_4(t,t')+\cO(J^5)\label{expC}\\\label{expR}
 R(t,t')&=\frac{1}{2}\lan\chi(t,t')\cos\bte(t,t')\ran=R_0(t,t')+J R_1(t,t')+J^2 R_2(t,t')
               +J^3 R_3(t,t')+J^4 R_4(t,t')+\cO(J^5).
\end{align}
Plugging these expansions in~\eqref{LEsc1} a hierarchy of equations for the $\te_n(t)$ results:
\begin{align}\label{eq00}
 \dte_0(t) &= \om +h(t)\\\label{eq01}
 \dte_1(t) &= \ph(t)\\\label{eq02}
 \dte_2(t) &= \psi(t)\te_1(t)-\int_0^t\!\!dt'\, R_0(t,t')\sin\bte_0(t,t')\\\label{eq03}
 \dte_3(t) &= \psi(t)\te_2(t)-\frac{1}{2}\ph(t)\te_1^2(t)
     -\int_0^t\!\!dt'\, \Big[R_0(t,t')\cos\bte_0(t,t')\,\bte_1(t,t')     +R_1(t,t')\sin\bte_0(t,t')\Big]\\\nonumber
 \dte_4(t) &= \psi(t)\te_3(t)-\ph(t)\te_1(t)\te_2(t)-\frac{1}{6}\psi(t)\te_1^3(t)
  -\int_0^t\!\!dt'\,\Big[R_0(t,t')\cos\bte_0(t,t')\,\bte_2(t,t')\\\label{eq04}
    &\qquad     -\frac{1}{2}R_0(t,t')\sin\bte_0(t,t')\bte_1^2(t,t')
         +R_1(t,t')\cos\bte_0(t,t')\, \bte_1(t,t')+R_2(t,t')\sin\bte_0(t,t')\Big].
\end{align}
From~\eqref{eq00} we immediately get the lowest order result
\begin{equation}\label{res0}
 \te_0(t)=\om t +\int_0^t\!\! dt'\, h(t')+\theta^{(0)},
\end{equation}
which when used in~\eqref{expC} with $h(t)\equiv 0$ yields
\begin{equation}\label{resC0}
 C_0(t,t')=\frac{1}{2}\lan\cos\om(t-t')\ran_{\om,\te^{(0)}}=\frac{1}{2}\,e^{-\frac{1}{2}(t-t')^2}.
\end{equation} 
Here the $\te^{(0)}$-average was trivial since no dependence on $\te^{(0)}$ remained, and the $\om$-average just involved a Gaussian integral from~\eqref{Pom}. Note that $C_0$ only depends on $\tau:=t-t'$.

Moreover, 
\begin{equation}\label{reschi0}
 \chi_0(t,t'):=\frac{\delta \te_0(t)}{\delta h(t')}=\kappa(t-t'):=\begin{cases}
                                                         1 &  t>t'\\
                                                         0 & \mathrm{else}
                                                        \end{cases}\;,
\end{equation}
where the Heaviside-function was denoted by $\kappa(t)$ in order to avoid confusion with the various $\te$ already around. From~\eqref{expR} now follows 
\begin{equation}\label{resR0}
 R_0(t,t')=\frac{1}{2}\lan\chi_0\cos\om(t-t')\ran_{\om,\te^{(0)}}=\kappa(t-t')\,C_0(t,t')
          =\kappa(t-t')\,\frac{1}{2}\,e^{-\frac{1}{2}(t-t')^2}.
\end{equation} 
This completes the leading order analysis. 

The first order terms, $C_1$ and $R_1$, each involve a single noise term $\xi_i$ and therefore average to zero, cf.~\eqref{eq01} and~\eqref{defphpsi}. To second order we have  
\begin{equation}\label{C2}
 C_2 = -\frac{1}{2}\Big[\lan\sin\bte_0\,\bte_2\ran+\frac{1}{2}\lan\cos\bte_0\,\bte_1^2\ran\Big],
\end{equation} 
where the average is over $\om$, $\te^{(0)}$ and the $\xi_i$ with correlations described by $C_0(t,t')$. Now 
\begin{align}\nonumber
 \lan\bte_1^2(t,t')\ran_\xi&=\int_{t'}^t\!\!dz_1\!\!\int_{t'}^t\!\!dz_2\,\lan\dte_1(z_1)\dte_1(z_2)\ran_\xi
   =\int_{t'}^t\!\!dz_1\!\!\int_{t'}^t\!\!dz_2\,\lan\ph(z_1)\ph(z_2)\ran_\xi
   =\int_0^\tau\!\!dz_1\!\!\int_0^\tau\!\!dz_2\,C_0(z_1-z_2)\cos\om(z_1-z_2)\\
  &=2\int_0^\tau\!\!du\,(\tau-u)\,C_0(u)\cos\om u
\end{align}
and 
\begin{align}\nonumber
 \lan\bte_2(t,t')\ran_\xi&=\int_{t'}^t\!\!dz_1\, \lan\dte_2(z_1)\ran_\xi
   =\int_{t'}^t\!\!dz_1\!\!\int_0^{z_1}\!\!\!\!dz_2\,
        \big[\lan\psi(z_1)\ph(z_2)\ran_\xi-R_0(z_1,z_2)\sin\bte_0(z_1,z_2)\big]\\\label{rest2}
   &= -2\int_0^\tau\!\!dz_1\!\!\int_{-t'}^{z_1}\!\!dz_2\,C_0(z_1-z_2)\sin\om(z_1-z_2)
    \to -2\tau\,\int^{\infty}_0\!\!du\,C_0(u)\sin\om u
\end{align}
where the limit $t,t'\to\infty$ with fixed $\tau=t-t'$ was taken. Therefore 
\begin{equation}\label{c2av}
 C_2(\tau)=\left\lan\tau\sin\om\tau\,\int^{\infty}_0\!\!du\,C_0(u)\sin\om u
              -\cos\om\tau\,\int_0^\tau\!\!du\,(\tau-u)\,C_0(u)\cos\om u\right\ran_{\om,\te^{(0)}}.
\end{equation} 
Again no dependence on the initial condition $\te^{(0)}$ remains. The average over $\om$ is accomplished by using
\begin{equation}\label{omav}
 \lan\sin\om\tau\,\sin\om u\ran_\om=e^{-\frac{1}{2}(\tau^2+u^2)}\sinh u\tau \spa{and}
 \lan\cos\om\tau\,\cos\om u\ran_\om=e^{-\frac{1}{2}(\tau^2+u^2)}\cosh u\tau.
\end{equation}
Performing finally the $u$-integrals we end up with
\begin{equation}\label{resC2erf}
C_2(\tau)=\frac{3\tau\sqrt{\pi}}{32}\,e^{-\frac{1}{4}\tau^2}\Big[
   3\,\erf\left(\frac{\tau}{2}\right)-\erf\left(\frac{3\tau}{2}\right)\Big] 
  + \frac{1}{16}\,e^{-\frac{1}{2}\tau^2}\Big[1-e^{-2\tau^2}\Big].
\end{equation} 
\begin{figure}[ht]
 \includegraphics[width=.35\textwidth]{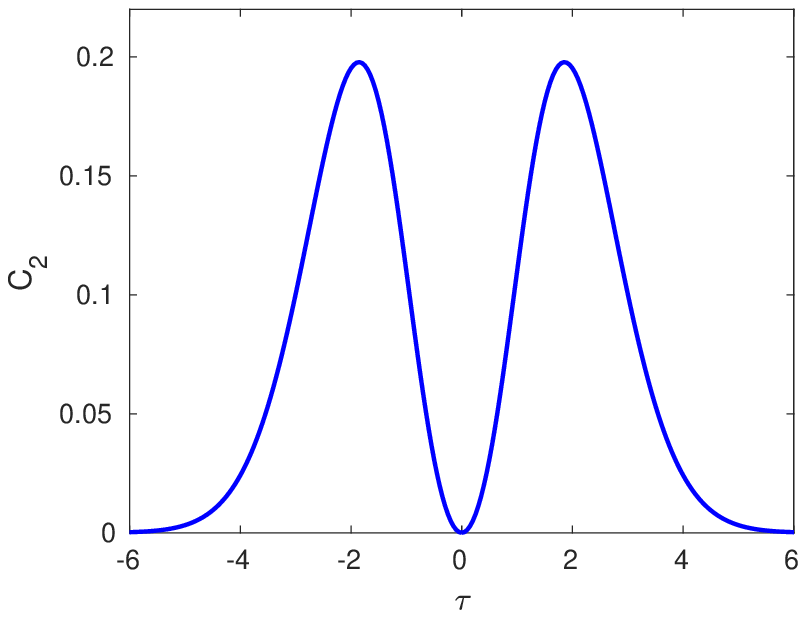}\hspace{10ex}
 \includegraphics[width=.35\textwidth]{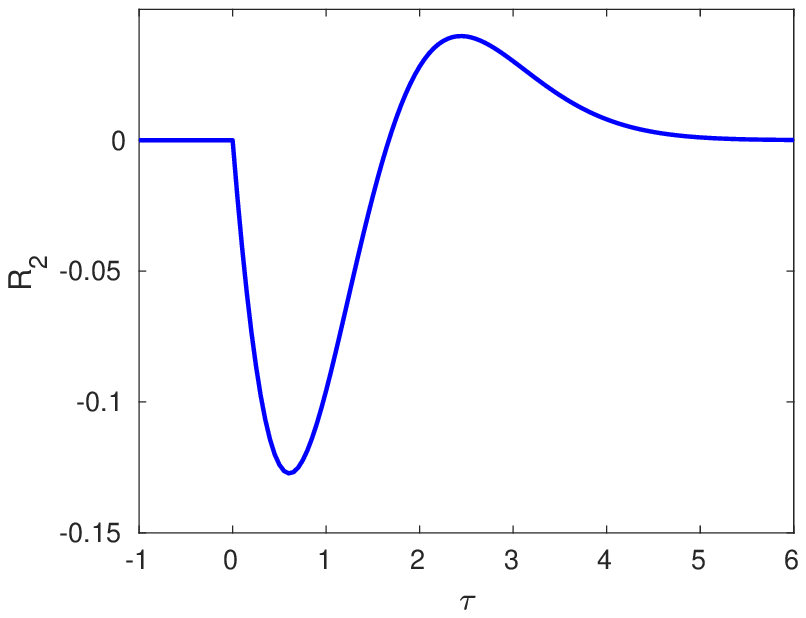}
 \caption{$C_2$ (left) as given by~\eqref{resC2erf} and $R_2$ (right) as given by~\eqref{resR2erf} as functions of $\tau$.}
 \label{fig1}
\end{figure}
To find the second order result for the response function we first need $\chi_1$ and $\chi_2$ that follow from~\eqref{eq00}-\eqref{eq02}. The contribution to $R_2$ involving $\chi_1$ again averages to zero
and we are left with 
\begin{equation}\label{res0R2} 
R_2(t,t')=\kappa(t-t') C_2(t,t')+\frac{1}{2}\lan \chi_2\,\cos\bte_0\ran.
\end{equation} 
For $\chi_2$ we get 
\begin{equation*}
 \lan\chi_2(t,t')\ran_{\xi}=\frac{\delta \lan\te_2(t)\ran_{\xi}}{\delta h(t')}
 =-2\,\kappa(t-t')\!\!\int_{t'}^t\!\!dz_1\!\!\int_0^{t'}\!\!\!\!dz_2\,C_0(z_1-z_2)\cos\om(z_1-z_2)
 \to -2\kappa(\tau)\!\!\int_0^\tau\!\!\!dz_1\!\!\int_{-\infty}^0\!\!\!\!\!\!\!dz_2\,C_0(z_1-z_2)\cos\om(z_1-z_2).
\end{equation*}
Multiplying with $\cos\om\tau$ and using again~\eqref{omav} we arrive at the final expression
\begin{equation}\label{resR2erf}
 R_2(\tau)=\kappa(\tau)\left(\frac{\tau\sqrt{\pi}}{32}\,e^{-\frac{1}{4}\tau^2}
   \Big[7\,\erf\left(\frac{\tau}{2}\right) + 3\,\erf\left(\frac{3\tau}{2}\right)-8\Big] - 
    \frac{1}{16}\,e^{-\frac{1}{2}\tau^2}\Big[1-e^{-2\tau^2}\Big]\right).
\end{equation} 
The second order results are shown in Fig.~\ref{fig1}. 

\begin{figure}[ht]
 \includegraphics[width=.35\textwidth]{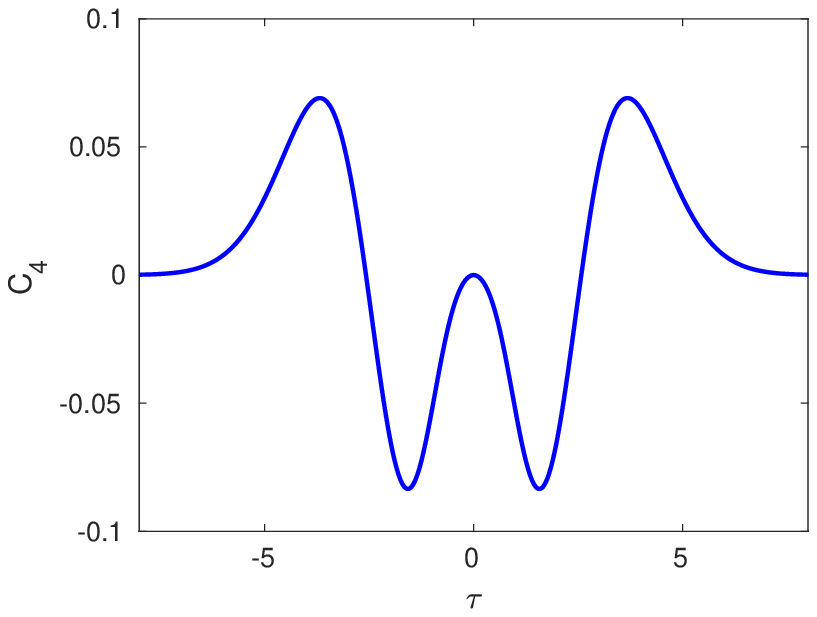}\hspace{10ex}
 \includegraphics[width=.35\textwidth]{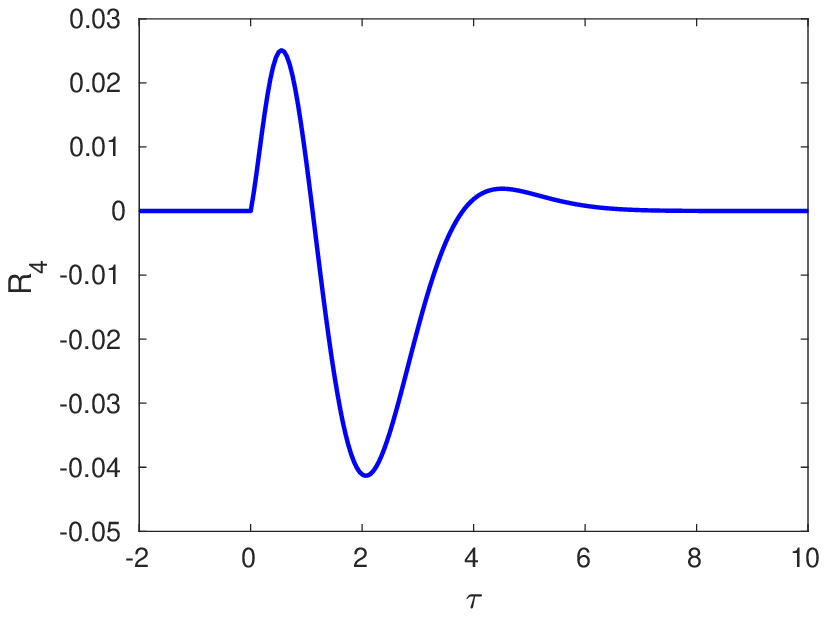}
 \caption{$C_4$ (left) and $R_4$ (right) as functions of $\tau$.}
 \label{fig2}
\end{figure}

The third order expressions, $C_3$ and $R_3$, again involve an odd number of noise terms $\xi_i$ and average to zero. The calculation of the fourth results requires the analysis of quite a few terms similar to but more involved than  those discussed above. Additional terms arise when the expansion~\eqref{expC} is used in~\eqref{statphpsi}. Nevertheless, the analysis is still feasible. In the final expressions for both $C_4$ and $R_4$ one integral has to be done numerically. Our results for $C_4(\tau)$ and $R_4(\tau)$, respectively, are shown in Fig.~\ref{fig2}. The combined perturbative results yield good approximations for correlation and response functions roughly up to $J\simeq 0.8$, cf. also Fig.~1 of the main text.


\section{Order parameter for the new oscillator}

From eq.~(15) of the main text the order parameter for the new oscillator is given by 
\begin{equation}\label{defP0}
 \Psi(t):=\frac{1}{J}\sum_j J_{0j}\, e^{i\tte_j(t)}.
\end{equation}
It is important to use $\tte_j$ instead of $\te_j$ in~\eqref{defP0} because the dynamics of the $\te_j$ is slightly modified by the addition of the new oscillator, cf.~\eqref{defdeltate}, and these modifications are correlated with $J_{0j}$.
Since $\delta\te_j\ll\te_j$ we may expand:
\begin{align}\nonumber
 \Psi(t)&=\frac{1}{J}\sum_j  J_{0j}\, e^{i\te_j(t)}(1+i\,\delta\te_j(t)+\dots)\\\nonumber
  &=\frac{1}{J}\sum_j J_{0j}\, e^{i\te_j(t)}+i\frac{1}{J}\sum_j J_{0j}\, e^{i\te_j(t)} \sum_l\int_0^t\!\! dt'\,\chi_{jl}(t,t') J_{l0}\sin\big(\te(t')-\te_l(t')\big)\\\nonumber
  &=\frac{1}{J}\sum_j J_{0j}\, e^{i\te_j(t)}
      +\frac{1}{2J}\sum_{j,l} \int_0^t\!\! dt'\,J_{0j}J_{l0}\,\chi_{jl}(t,t')
          e^{i\te_j(t)} \left[e^{i(\te(t')-\te_l(t'))}
         -e^{-i(\te(t')-\te_l(t'))}\right]\\\label{decomppsi}
  &=\xi(t)+\zeta(t),
\end{align}
where 
\begin{equation*}
 \xi(t):=\xi_1(t)+i\xi_2(t)
\end{equation*}
with $\xi_1$ and $\xi_2$ as defined in \eqref{defxi} and
\begin{equation*}
 \zeta(t):=\frac{1}{2J}\int_0^t\!\! dt'\,\left[
    e^{i\te(t')}\sum_{j,l} J_{0j}J_{l0}\,\chi_{jl}(t,t')\,e^{i(\te_j(t)-\te_l(t'))}
   -e^{-i\te(t')}\sum_{j,l} J_{0j}J_{l0}\,\chi_{jl}(t,t')\,e^{i(\te_j(t)+\te_l(t'))}\right].
\end{equation*} 
Also $\zeta(t)$ is a random quantity. In view of the discussion between eqs.~\eqref{statR2a} and~\eqref{defR2}, however, we realize that its fluctuations induced by the randomness in the $J_{0i}$ and the $\te_i(t)$ with $i\geq1$ can be neglected. We may therefore write
\begin{equation}\label{defzeta}
 \zeta(t)=J\int_0^t\!\! dt'\,R(t,t')\,e^{i\te(t')}=:\zeta_1(t)+i\zeta_2(t)=:\rho(t)\,e^{i\ph(t)},
\end{equation} 
with
\begin{equation}\label{defrho}
 \rho(t)=J \left|\int_0^t\!\! dt'\,R(t,t')\,e^{i\te(t')}\right|.
\end{equation} 
Note, that $\zeta(t)$ and $\rho(t)$ are still random functions due to their dependence on $\om,\, \te^{(0)}$ and $\xi_i$. The probability distribution of
\begin{equation*}
 r(t):=\left|\Psi(t)\right|=\sqrt{(\xi_1+\zeta_1)^2+(\xi_2+\zeta_2)^2}
\end{equation*}
is hence given by 
\begin{equation}\label{defpofrho}
 P(r(t)):=\left\lan \int\!\!\cD\xi_1(\cdot)\cD\xi_2(\cdot)\,
     \delta\big(r(t)-\left|\Psi(t)\right|\big)\right\ran_{\om,\te^{(0)}}.
\end{equation}
Actually, we are more interested in 
\begin{equation}\label{deftP}
 \tP(r):=\frac{P(r)}{2\pi r},
\end{equation} 
because it is this distribution that shows the volcano transition. 
\begin{figure}[t]
 \includegraphics[width=.3\textwidth]{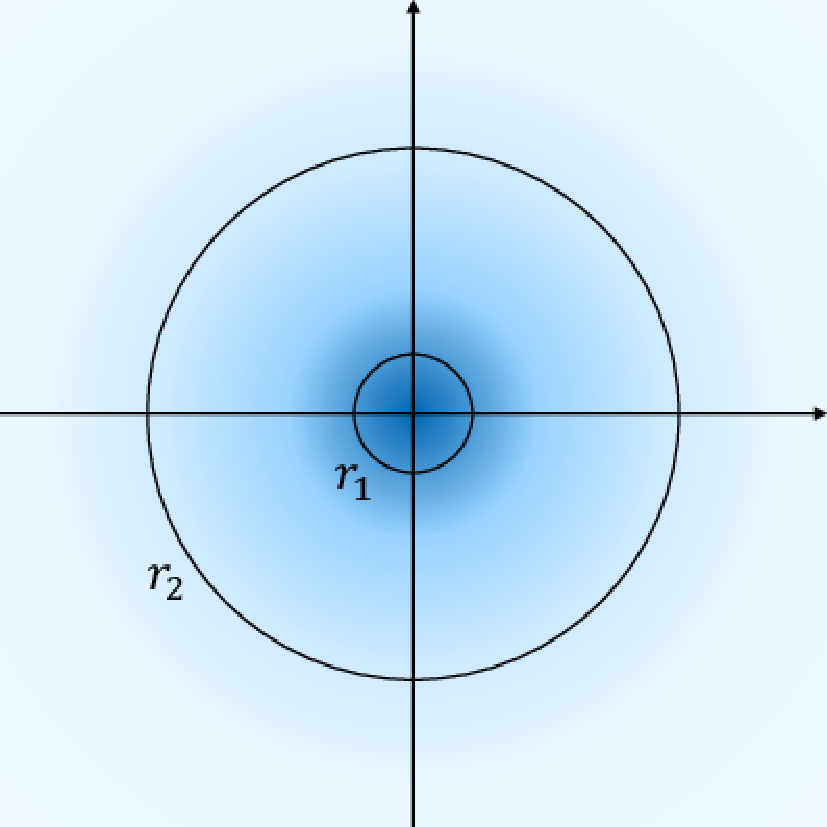}\hspace{8ex}
 \includegraphics[width=.3\textwidth]{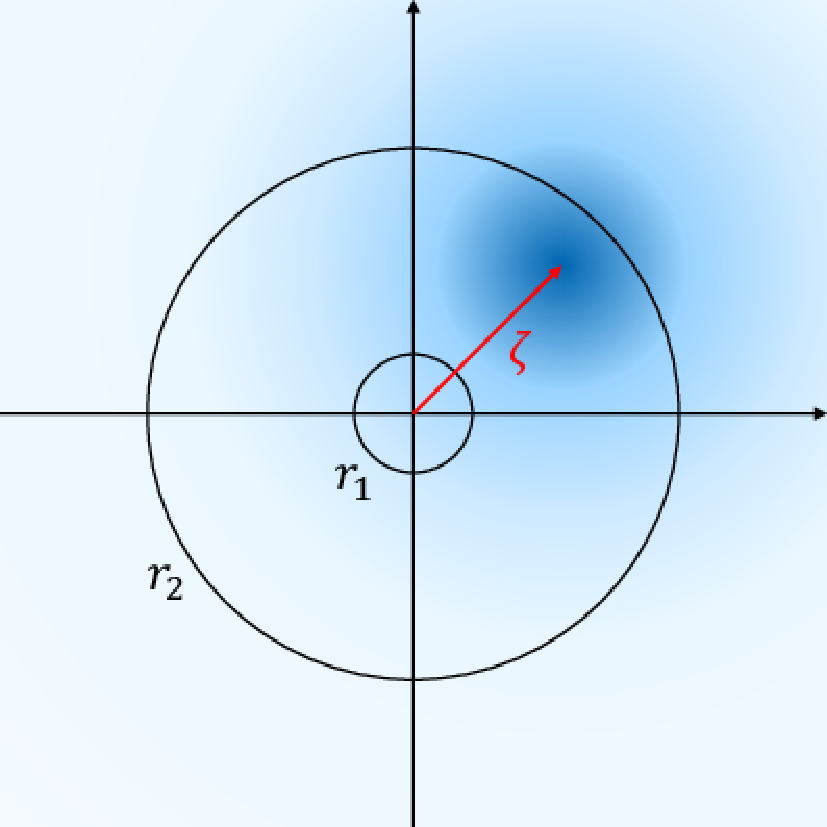}
 \caption{Mechanism of the volcano transition. Shown are color coded plots of $P_\Psi$ over the complex $\Psi$-plane for $J=0$ (left) and for sufficiently large $J>0$ (right). The function $\tP(r)$ defined in~\eqref{deftP} is the average of $P_\Psi$ over a circle with radius $r$ centered at the origin. Two such circles are explicitly shown. By inspection one infers $\tP(r_1)>\tP(r_2)$ for the left subfigure, but $\tP(r_1)<\tP(r_2)$ for the right one.}
 \label{fig4}
\end{figure}
The decomposition~\eqref{decomppsi} of $\Psi$ now offers a clear intuitive understanding of this transition. In Fig.~\ref{fig4} the shading displays the probability distribution $P_\Psi$ of the complex order parameter $\Psi$. The left subfigure is for $J=0$. Then $\zeta=0$ and by~\eqref{decomppsi} $\Psi$ inherits the Gaussian distribution centered at the origin from the noise $\xi$. Moreover, it is clear from \eqref{deftP} that $\tP(r)$ is nothing but the average of $P_\Psi$ along a circle with radius $r$ centered at the origin of the complex $\Psi$-plane. Consequently, for $J=0$, we always have $\tP(r_1)>\tP(r_2)$ if  $r_1<r_2$ and the maximum of $\tP$ must be at $r=0$. With increasing $J$, however, $\zeta$ becomes nonzero implying that the center of $P_\Psi$ gets shifted away from $\Psi=0$ as shown in the right subfigure of Fig.~\ref{fig4}. For sufficiently large $|\zeta|$ the average of $P_\Psi$ over the larger circle may now be {\em larger} than the one over the smaller circle. Accordingly, the maximum of $\tP$ got shifted to some value $r>0$. 

\begin{figure}[ht]
 \includegraphics[width=.4\textwidth]{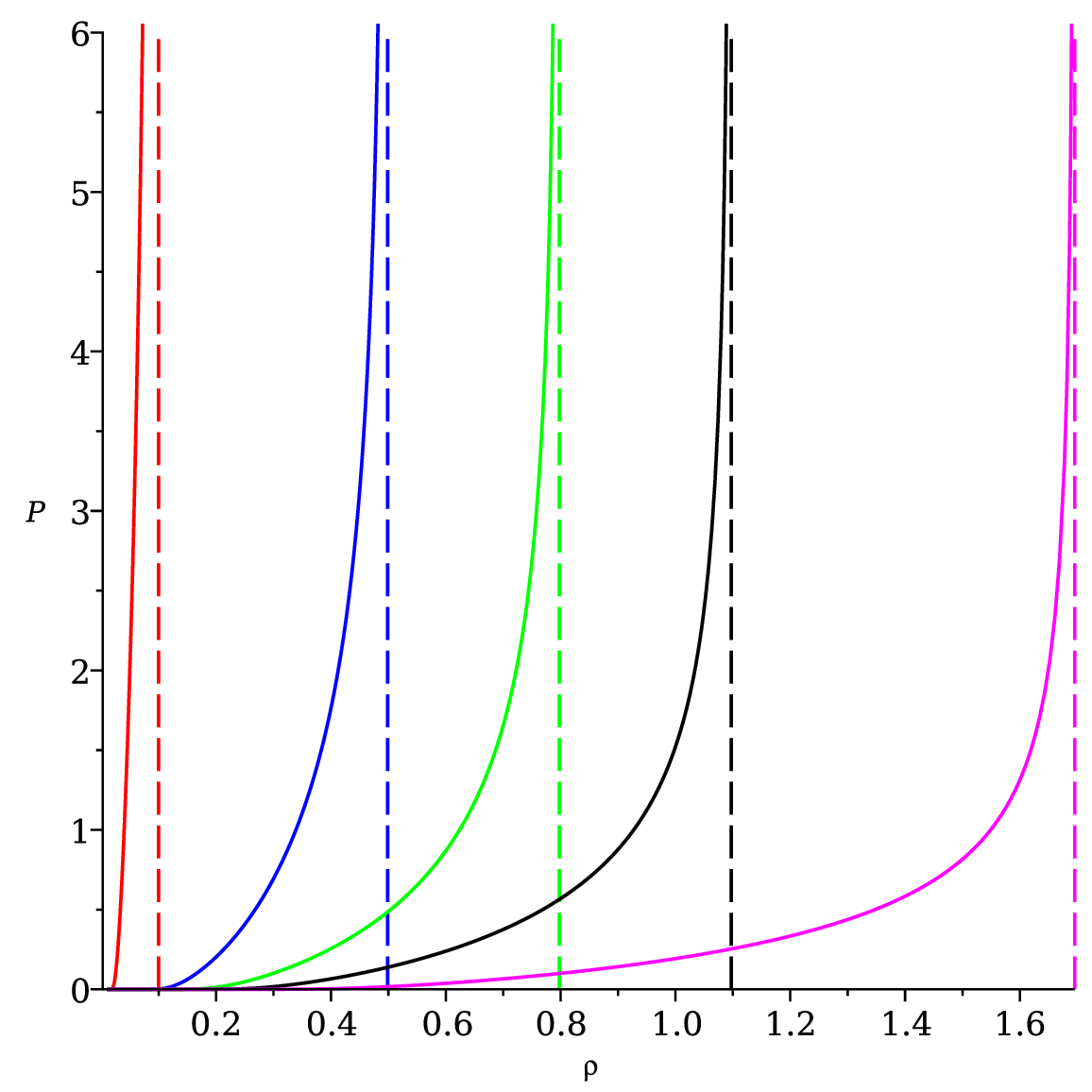}
 \caption{Distribution of $\rho$ in lowest order perturbation theory for the five values of $J=0.16, 0.80, 1.27, 1.75, 2.70$. The dashed lines indicate the upper bounds given by~\eqref{defrhoup}.}
 \label{fig3}
\end{figure}

To substantiate this intuitive understanding a more explicit expression for $\tP(r)$ were desirable. However, from~\eqref{defzeta} we see that $\zeta(t)$ depends via $\te(t')$ on $\xi(t')$ for all $t'\leq t$. It seems impossible to account for these correlations exactly and we have to resort to approximations. To this end it is instructive to consider the case of small $J$ and to use the lowest order perturbation results 
\begin{equation*}
 \te(t)=\te_0(t)=\om t +\te^{(0)}\spa{and} R(\tau)=R_0(\tau)=\kappa(\tau)\,\frac{1}{2}\,e^{-\frac{\tau^2}{2}}.
\end{equation*}
Eq.~\eqref{defrho} then simplifies to
\begin{align}\nonumber
 \rho\simeq\rho_0&= \frac{J}{2}\left|e^{i\te^{(0)}}\int_0^t\!\! dt'\,e^{-\frac{(t-t')^2}{2}}\,e^{i\om t'}\right|
   =\frac{J}{2}\left|e^{i\om t}\int_0^t\!\! d\tau\,e^{-\frac{\tau^2}{2}}\,e^{-i\om \tau}\right|\\\nonumber
  &\to\frac{J}{2}\left|\int_0^\infty\!\! d\tau\,e^{-\frac{\tau^2}{2}}\,\big(\cos\om \tau-i\sin\om \tau\big)\right|
   =\frac{J}{2}\sqrt{\frac{\pi}{2}}\,e^{-\frac{\om^2}{2}}\left|1-\erf\left(\frac{i\om}{\sqrt{2}}\right)\right|\\
  &=\frac{J}{2}\sqrt{\frac{\pi}{2}}\,e^{-\frac{\om^2}{2}}\sqrt{1-\erf^2\left(\frac{i\om}{\sqrt{2}}\right)}
    =:J f(\om).\label{h4}
\end{align}
Here the dependence on $\te^{(0)}$ dropped out and the limit $t\to\infty$ was taken. From~\eqref{h4} we obtain
\begin{equation*}
 P(\rho_0)=\frac{P_\om(\om(\rho_0))}{J|f'(\om(\rho_0))|}.
\end{equation*}
Inverting $f(\om)$ numerically and using~\eqref{Pom} we determine $P(\rho_0)$. Fig.~\eqref{fig3} shows the corresponding results for five values of $J$ that were also originally considered by Daido.

As is clearly seen the most probable values of $\rho_0$ are those approaching the upper bound 
\begin{equation}\label{defrhoup}
 \rho_u:=J\int_0^t\!\! dt'\,\left|R(t,t')\,e^{i\te(t')}\right|=J\int_0^t\!\! dt'\,|R(t,t')|\, .
\end{equation} 
Building on the result of lowest order perturbation theory this reasoning remains restricted to small values of $J$. 
However, since we expect an increasing tendency for $\te$ to move with $\dte<\om$ or to even freeze for larger values of $J$ the approximation $\rho\simeq\rho_u$ is presumably also quite accurate in this regime. This assumption is corroborated by numerical results for $P(\rho)$.

Employing the approximation~\eqref{defrhoup} in~\eqref{defpofrho} no dependence on either $\om$ nor $\te^{(0)}$ remains  and there are no longer non-local correlations between $\zeta(t)$ and $\xi(t')$. We may hence replace the path integrals by simple integrals over $\xi_i(t)$. Using~\eqref{corxi} for $t=t'$ we then find
\begin{align*}
 P(r)\simeq P_u(r)&=\int\!\!d\xi_1d\xi_2\,P(\xi_1,\xi_2)\,
         \delta\left(r-\sqrt{(\xi_1+\zeta_1)^2+(\xi_2+\zeta_2)^2}\right)\\
 &=\int\!\!\frac{d\xi_1d\xi_2}{\pi}\,e^{-\xi_1^2-\xi_2^2}\,
         \delta\left(r-\sqrt{(\xi_1+\zeta_1)^2+(\xi_2+\zeta_2)^2}\right)
 =\int\!\! \frac{d\eta_1, d\eta_2}{\pi}\,e^{-\eta_1^2-\eta_2^2}\;
      \delta\left(r-\sqrt{\eta_1^2+(\eta_2+\rho_u)^2}\right)\\
 &=\frac{r}{\pi}\int\!\! dx \,d\eta_1\,e^{-\eta_1^2-(x-\rho_u)^2}\;\frac{1}{\sqrt{r^2-x^2}}
      \left[\delta\Big(\eta_1+\sqrt{r^2-x^2}\Big)+\delta\Big(\eta_1-\sqrt{r^2-x^2}\Big)\right]\\
 &=\frac{2r}{\pi}\,e^{-r^2-\rho_u^2}\int_{-r}^r\frac{dx}{\sqrt{r^2-x^2}} \,e^{2 x\rho_u}
  =2r\,e^{-r^2-\rho_u^2}\,I_0(2r\rho_u).
\end{align*}
This implies 
\begin{equation*}
 \tP_u(r):=\frac{P_u(r)}{2\pi r}=\frac{1}{\pi}\,e^{-r^2-\rho_u^2}\,I_0(2r\rho_u)
\end{equation*}
which coincides with eq.~(21) in the main text. 


\section{Numerics}

The numerical determination of the self-consistent correlation and response function as defined by the system of eqs.~(4,~5,~8-10) of the main text is a two-level procedure. In an inner loop, for given $C(t,t')$ and $R(t,t')$, a family of $K=10^3..10^4$ independent pairs of solutions $\te_k(t),\chi_k(t,t'),\;k=1,...,K$ of eqs.~\eqref{LEsc1} and~\eqref{eomchi} is generated. To this end first $K$ values $\om_k$ and $\te_k^{(0)}$ are drawn from \eqref{Pom} and \eqref{Pte0}, respectively. Then, for each $k$ two noise histories $\xi_{k1}(t)$ and $\xi_{k2}(t)$ are generated according to \eqref{avxi}-\eqref{corxi} from a Cholesky decomposition of the matrix $C(t,t')$. Eqs.~\eqref{LEsc1} and~\eqref{eomchi} are then integrated using a standard ODE-solver with stepsize $\Delta t=0.05$ up to $M=400$ time steps. Finally, from the trajectories $\te_k(t)$ and $\chi_k(t,t')$ generated new estimates for $C(t,t')$ and $R(t,t')$ are determined via
\begin{equation*}
 C_\mathrm{new}(t,t')=\frac{1}{2K}\sum_k \cos\big(\te_k(t)-\te_k(t')\big) \spa{and}
 R_\mathrm{new}(t,t')=\frac{1}{2K}\sum_k \chi_k(t,t')\cos\big(\te_k(t)-\te_k(t')\big).
\end{equation*}

In an outer loop the correlation and response function are iterated to their self-consistent forms. As initial guess we start with the lowest order perturbation results $C(t,t')=C_0(t-t')$ and $R(t,t')=R_0(t-t')$ as given by \eqref{resC0} and \eqref{resR0}, respectively. After one iteration of the inner loop we use the functions $C_\mathrm{new}$ and $R_\mathrm{new}$ for a soft update 
\begin{equation}\label{reinj}
 C \longrightarrow (1-a)C+a C_\mathrm{new},\quad
 R \longrightarrow (1-a)R+a R_\mathrm{new}.
\end{equation}
A suitable value for the reinjection parameter is $a=0.3$. The updated forms of $C$ and $R$ are then used for the second iteration of the inner loop and so on.

The stopping criterion of the whole procedure has to cope with the statistical nature of $C$ and $R$. Even if these functions have already reached their self-consistent form a new run of the inner loop will produce values for $C_\mathrm{new}$ and $R_\mathrm{new}$ different from $C$ and $R$ due to the randomness in $\om, \te^{(0)}$ and the noise sources $\xi_1$ and $\xi_2$. The problem is hence to disentangle these stochastic fluctuations from the average convergence of the outer loop. Assume we have obtained functions $C_\mathrm{trial}$ and $R_\mathrm{trial}$ and we would like to test whether they are already self-consistent. We generate two groups containing each five matrices $C_n(t,t'), n=1,..,5$ (equivalently for $R$). The $C_n$ of the first group result from five iterations of the inner loop each starting with $C_\mathrm{trial}$ and $R_\mathrm{trial}$. This group hence estimates the statistical fluctuations of $C$. For the second group we also start the first iteration of the inner loop with $C_\mathrm{trial}$ and $R_\mathrm{trial}$ but then continue to adapt $C$ and $R$ according to \eqref{reinj}. In this way the tendency to further improvement is tested. Using the Frobenius norm we compare the average differences between successive matrices in both groups. The outer loop is stopped when the differences in the second group are less than 5\% larger than those in the first one. 

Once the self-consistent forms of $C$ and $R$ are found much longer trajectories $\te(t)$ with up to $2\cdot 10^5$ time steps are generated to collect the data shown in the figures. 

The numerical solution of the original $N$-oscillator network is again done with a standard ODE-solver with step size $\Delta t=0.05$ and up to $2\cdot 10^5$ time steps.